\documentclass[]{aastex}

\usepackage{emulateapj5}
\usepackage{onecolfloat}
\usepackage{graphicx} 
\usepackage{fancyheadings} 
\usepackage{ulem}
\usepackage{rotating}
\usepackage{lscape}

\newcommand{\bdt}{BD+28$^\circ$4211}
\newcommand{\nobs}{seven}

\newcommand{\lya}{Ly$\alpha$}

\newcommand{\kms}{km~s$^{-1}$ }
\newcommand{\cm}[1]{\, {\rm cm^{#1}}}
\newcommand{\N}[1]{{N({\rm #1})}}

\newcommand{\sci}[1]{{\rm \; \times \; 10^{#1}}}

\begin{document}

\twocolumn[%
\submitted{Accepted to the Astrophysical Journal Letters: December 28, 2004}
\title{Evidence for Correlated Titanium and Deuterium Depletion
in the Galactic ISM}

\author{ Jason X. Prochaska\altaffilmark{1,2},
Todd M. Tripp\altaffilmark{3},
J. Christopher Howk\altaffilmark{2,4}}
\begin{abstract} 

Current measurements indicate that the deuterium abundance in diffuse
interstellar gas varies spatially by a factor of $\sim$4 among
sightlines extending beyond the Local Bubble.  
One plausible explanation for the scatter is the variable depletion
of D onto dust grains.  To test this scenario, we have obtained
high signal-to-noise, high resolution profiles of the refractory ion
\ion{Ti}{2} along seven Galactic sightlines with D/H ranging
from 0.65 to $2.1 \sci{-5}$.  These measurements, acquired with the
recently upgraded Keck/HIRES spectrometer, indicate a correlation
between Ti/H and D/H at the $> 95\%$ c.l.
Therefore, our observations support the interpretation that D/H scatter
is associated with differential depletion.  We note, however, that 
Ti/H values taken from the literature do not uniformly show the
correlation.  
Finally, we identify significant component-to-component variations 
in the depletion levels
among individual sightlines and discuss complications arising from this
behavior.  
  
\keywords{cosmology: observations, ISM: evolution, ISM: abundances, 
galaxy: abundances}

\end{abstract}
]

\altaffiltext{1}{UCO/Lick Observatory; University of California, Santa
  Cruz; Santa Cruz, CA 95064, {\tt xavier@ucolick.org}}
\altaffiltext{2}{Visiting Astronomer, W. M. Keck Telescope.
The Keck Observatory is a joint facility of the University
of California and the California Institute of Technology.}
\altaffiltext{3}{Department of Astronomy;
University of Massachusetts; 710 N. Pleasant St.; Amherst, MA 01003-9305}
\altaffiltext{4}{Department of Physics, and Center for Astrophysics and 
Space Sciences, University of California, San Diego, C--0424, La Jolla, 
CA 92093-0424}

\pagestyle{fancyplain}
\lhead[\fancyplain{}{\thepage}]{\fancyplain{}{PROCHASKA, TRIPP, \& HOWK}}
\rhead[\fancyplain{}{Evidence for Correlated Titanium and Deuterium Depletion
in the Galactic ISM}]{\fancyplain{}{\thepage}}
\setlength{\headrulewidth=0pt}
\cfoot{}

\section{INTRODUCTION}
\label{sec-intro}

In standard big bang nucleosynthesis (BBN), 
deuterium is the most sensitive baryometer of the light elements
\citep{schramm98}. 
Measurements of its primordial value in quasar
absorption line systems have placed tight
constraints on the baryonic mass density $\Omega_b$ 
\citep{burles98,omeara01,kirkman03}.
The inferred $\Omega_b$ value is in impressive agreement with the
value derived from CMB measurements of the WMAP \citep[e.g.][]{spergel03} 
and other microwave experiments lending
confidence to both the deuterium measurements and BBN theory.
In addition to its cosmological significance, deuterium is
an important tracer of chemical evolution.
Deuterium is astrated within stellar cores, and there
are no known means of producing significant amounts of D other than BBN.
The evolution of D/H, therefore, tracks the global history
of the gas astration within a galaxy \citep[e.g][]{chiappini02}.

It is in this context that measurements of D/H 
within the Milky Way have impact:
(1) the upper bound to the Galactic D/H value
sets a lower limit to the primordial D/H value; and
(2) a comparison of Galactic D/H with the 
primordial D/H value describes the chemical
evolution history of the Galaxy.
Since the first measurements of Galactic D/H with {\it Copernicus}
\citep{rogerson73},
surveys for Galactic D/H have been pursued on each succeeding 
UV observatory bearing a high resolution spectrometer 
\citep[e.g.][]{linsky95,jenkins99,hoopes03,wood04}.
The most remarkable conclusion of these efforts is that beyond
the Local Bubble the D/H values have significant scatter;
D/H ranges from 5 to $20 \sci{-6}$.
This dispersion is larger than the estimated statistical error and
is unlikely to be systematic error associated with the
data analysis (see references in Table~\ref{tab:summ} for
detailed and thorough analyses of the measurement uncertainties).  
At present, it seems that the 
D/H ratio has intrinsic scatter within the Milky Way.

Recently, observations with the Interstellar Medium Absorption
Profile Spectrograph (IMAPS) and the {\it FUSE Observatory} 
have extended D/H measurements to large distances from the 
Sun and correspondingly
higher $\N{HI}$ values.  Although only a handful of measurements
exist to date, \cite{wood04} and others have noted that the
few sightlines that probe the greatest distances ($>500$pc)
have a central value and dispersion that
are significantly smaller than sightlines at intermediate distances
(20\,pc~$< d < 500$\,pc).
One possible interpretation of these trends is that D is
significantly depleted onto dust grains \citep[e.g.][]{jura82}.
\cite{draine04a,draine04b} has examined the principal
mechanisms of D adsorption and grain destruction and argues
that it is at least plausible that D would be preferentially depleted.
He proposed testing the depletion hypothesis by comparing the
D/H values with abundances of refractory elements like Fe, Ni, and
Ti (i.e.\ species that are highly prone to depletion onto grains)
along the same sightlines.

\begin{table*}[ht]
\begin{center}
\caption{{\sc SUMMARY OF OBSERVATIONS\label{tab:summ}}}
\begin{tabular}{cccccccc}
\tableline
\tableline
Target & $\epsilon$\,Ori &$\delta$\,Ori &HD\,191877 &HD\,195965 &
BD+28$^\circ$4211 &$\iota$\,Ori &Feige\,110 \\
\tableline
$B$& 1.71& 2.01& 6.22& 6.60&13.01& 2.53&13.01\\
Obs (UT)&08Sep2004&08Sep2004&08Sep2004&08Sep2004&06Oct2004&08Sep2004&09Sep2004\\
t$_{exp}$ (s)&   9&  13&1800&1200&1800&   9&1800\\
S/N$^a$&370&355&350&250&145&280&120\\
$\log[\N{HI}/\cm{-2}]$&$20.40 \pm 0.08$&$20.19 \pm 0.03$&$21.05 \pm 0.05$&$20.95 \pm 0.03$&$19.85 \pm 0.02$&$20.16 \pm 0.10$&$20.14 \pm 0.07$\\
D/H$^b$&$0.646 \pm 0.38$&$0.736 \pm 0.12$&$0.776 \pm 0.20$&$0.851 \pm 0.17$&$1.380 \pm 0.10$&$1.413 \pm 0.88$&$2.138 \pm 0.43$\\
Ref& 1& 2& 3& 3& 4& 1& 5\\
$\log[\N{Ti\,II~3230}]$&$<11.55$&$<11.46$&$12.26 \pm 0.03$&$11.97 \pm 0.05$&$<11.57$&$<11.67$&$<11.77$\\
$\log[\N{Ti\,II~3242}]$&$11.50 \pm 0.05$&$11.28 \pm 0.06$&$12.25 \pm 0.02$&$12.06 \pm 0.02$&$<10.99$&$11.20 \pm 0.07$&$11.69 \pm 0.05$\\
$\log[\N{Ti\,II~3384}]$&$11.37 \pm 0.04$&$11.09 \pm 0.06$&$12.23 \pm 0.02$&$12.00 \pm 0.02$&$11.08 \pm 0.08$&$11.34 \pm 0.03$&$11.54 \pm 0.04$\\
$\log[\N{Ti\,II}]^c$&$11.40 \pm 0.03$&$11.15 \pm 0.04$&$12.24 \pm 0.02$&$12.02 \pm 0.02$&$11.08 \pm 0.08$&$11.31 \pm 0.03$&$11.59 \pm 0.03$\\
$\log($Ti/H)&$-9.00 \pm 0.09$&$-9.05 \pm 0.05$&$-8.81 \pm 0.05$&$-8.93 \pm 0.03$&$-8.77 \pm 0.08$&$-8.85 \pm 0.10$&$-8.55 \pm 0.09$\\
\tableline
\end{tabular}
\end{center}
\tablenotetext{a}{Empirically measured per pixel at 3250\AA.}
\tablenotetext{b}{Units of 10$^{-5}$}
\tablenotetext{c}{Weighted mean}
 \tablerefs{Key to References -- 1: \cite{laurent79};
2: \cite{jenkins99};
3: \cite{hoopes03};
4: \cite{sonneborn02};
5: \cite{friedman02}}
\end{table*}

We have initiated a program
to obtain high signal-to-noise (S/N), high resolution observations
of \ion{Ti}{2} profiles for sightlines with accurate Galactic
D/H measurements.  Previous surveys of Ti have demonstrated
that it is highly refractory, presumably because of
its large condensation temperature \citep[e.g.][]{stokes78,lipman95}.
Therefore, a measurement of Ti/H in the ISM assesses
the depletion level along that sightline.  Of additional importance,
Ti$^+$ is the dominant ion in \ion{H}{1} regions. Unlike 
Na$^0$ and Ca$^+$ the ionization potential of Ti$^+$
is $\approx 1$\,Ryd and Ti$^+$ is predominantly 
shielded from ionizing photons.  Furthermore, 
Ti$^+$ has a charge exchange reaction rate with hydrogen that
is significantly greater than many other ions 
\citep[e.g.\ Fe$^+$,Si$^+$;][]{kingdon96}.
We expect is less sensitive to photoionization effects 
and \ion{Ti}{2} should trace 
the velocity profiles of \ion{H}{1} and \ion{D}{1} gas.
Therefore, high resolution \ion{Ti}{2} profiles are likely to be better
suited than, e.g., \ion{Fe}{2}, for constraining the fits of \ion{D}{1}
and assessing the likelihood of deuterium line saturation in 
high $\N{HI}$ sightlines.

In this paper, we report on our first set of \ion{Ti}{2} observations
of \nobs\ sightlines.  We measure the Ti$^+$ column densities
to assess the depletion levels along the sightlines and examine
correlations with the observed D/H values.  Finally,
this Letter establishes a public database for \ion{Ti}{2} 
measurements obtained by our 
group\footnote{http://www.ucolick.org/$\sim$xavier/TiII/index.html}.
The data presented here, and all future observations, will be archived
at this site, including the raw data and calibration frames.  
The data are freely available to other researchers.

\section{Observations and Reduction}

The \nobs\ sightlines presented here were observed on the nights 
of September 8 and 9 and October 6, 2004 UT with the recently
upgraded HIRES spectrometer \citep{vogt94} on the 10m Keck I telescope.
The spectrometer now contains a three CCD mosaic with $>90\%$
quantum efficiency at $\lambda < 3500$\AA.
Because the targets are bright, we obtained
the data primarily during twilight.
The first half of the first night 
was marred by poor observing conditions while the
remainder of the time was clear with typical
seeing of FWHM~$\approx 0.7''$.   The targets in 
September 2004 were observed through the B5 decker 
($0.86''$ width; FWHM~$\approx 6$\kms resolution; $3.5''$ slit length) 
while \bdt\ was observed through the B1 decker
($0.57''$ width; FWHM~$\approx 4.5$\kms resolution; $3.5''$ slit length). 
Table~\ref{tab:summ} summarizes the exposure times and data
quality of the sample.

At the time of publication, a data reduction pipeline for the
upgraded HIRES instrument was not available. 
Therefore, we reduced the two echelle orders containing
the \ion{Ti}{2} $\lambda 3230, 3242, 3384$ 
transitions with a set of in-house IDL routines.
These routines subtracted the bias,  interactively set a boxcar
aperture, interactively determined a region for scattered light
subtraction (sky background was minimal), and extracted a 1D
spectrum.  A variance array was calculated accounting for the
read noise and assuming Gaussian statistics. 
Wavelength calibration was carried out by fitting a 3rd order
polynomial to the arc lines identified in the 1D spectra 
extracted from a ThAr image using the same trace and aperture
as the science extraction.  The typical RMS deviation from the fit was
less than 0.1~pixels (i.e.\ $<0.003$\AA).
Although we achieved a formal S/N ratio $> 500$~pix$^{-1}$ for
the brightest targets, the scatter around the
stellar continuum significantly exceeds Gaussian statistics in
the brightest sources.  This is most likely due to noise in the
combined flat field or perhaps small errors in the traces of the
dispersed spectra.
Throughout the analysis,
we have augmented the variance arrays to match the
empirical scatter measured in the stellar continuum.

All of the individual exposures were rebinned to a common wavelength 
scale with $\lambda_0 = 3000$\AA\ and a pixel size of 1.4\kms\
after correcting to vacuum wavelengths and 
the heliocentric velocity reference frame.  
Multiple exposures of the same object were 
compared to identify cosmic rays and then coadded after scaling to 
a common flux and weighting by the square of the median signal-to-noise. 
Finally, we normalized
the data by fitting a high order polynomial to the unfluxed stellar
continuum.  Continuum normalization is generally the largest source of
uncertainty in the analysis, especially for 
the weakest transitions.

\begin{figure}[ht]
\begin{center}
\includegraphics[width=3.6in]{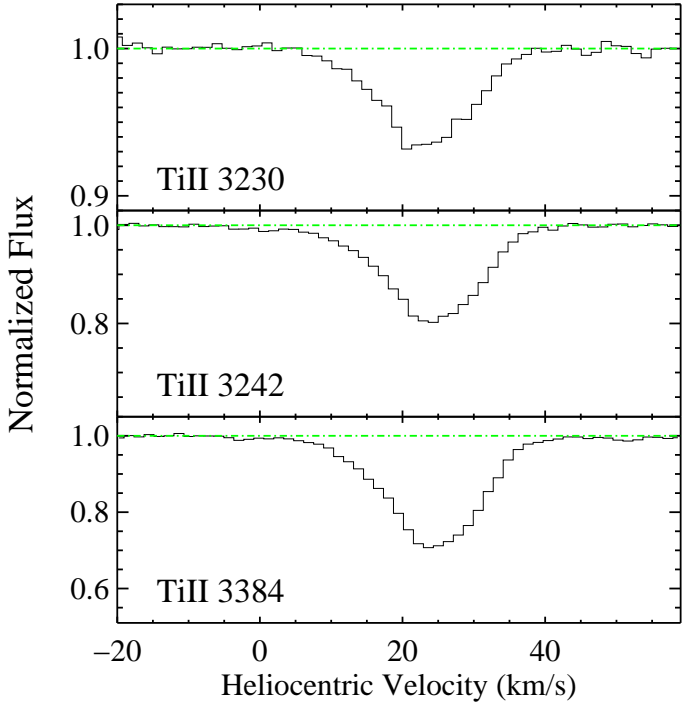}
\caption{Ti\,II profiles for the ISM sightline toward HD\,191877.
\label{fig:data}}
\end{center}
\end{figure}

\section{Analysis}

Figure~\ref{fig:data} presents a velocity plot for the HD~191877
sightline.  At the recorded S/N and resolution, it is straightforward
to measure the \ion{Ti}{2} ionic column densities. 
In the following, we assume that the majority of Ti$^+$ is in the 
$^{4}{\rm F}_{3/2}$ ground state.  From our observations, we measure
$\N{Ti\,II^* \lambda 3239} < 10^{10.6} \cm{-2}$ as a conservative
upper limit to the $^{4}{\rm F}_{5/2}$ state ($E \approx 135$\,K).
Provided the \ion{Ti}{2} profiles have small optical depth, we can
measure the \ion{Ti}{2} column density by (i) summing the 
EW and assume the weak limit for the curve-of-growth (the values
range from several to 30\,m\AA); 
(ii) integrating the line profile with the apparent optical depth
method \citep[AODM;][]{sav91,jenkins96}; 
and (iii) by fitting Voigt profiles to the data.
All three techniques yield similar results; we have mainly used the
AODM in this paper.
The results for the three transitions\footnote{Assuming
$f_{3230}=0.0687, f_{3242}=0.232, f_{3384}=0.358$ 
\citep{bizzarri93,pickering01,pickering02,morton03}.} for
all of the sightlines are listed in Table~\ref{tab:summ}.
For all measurements we report $1\sigma$ uncertainties and the upper 
limits correspond to $3\sigma$ limits.
The uncertainties include (in quadrature) statistical uncertainty
from Poisson noise and error due to continuum placement.
We estimate a 1$\sigma$ continuum error of 0.1$\%$ for all of the objects 
except Feige\,110 (0.2$\%$) which corresponds to 
an approximately 0.5\,m\AA\ error over the integrated profiles.
Finally, we adopt a minimum error of 0.02\,dex owing to systematic
uncertainties related to data reduction (e.g.\ flat fielding).

\begin{figure}[ht]
\begin{center}
\includegraphics[height=3.4in,angle=90]{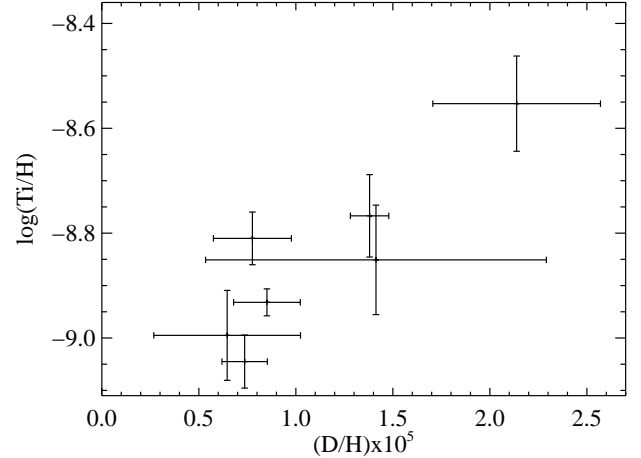}
\caption{$\log$(Ti/H) vs.\ D/H for the \nobs\ sightlines comprising
the current sample.  Error bars reflect $1\sigma$ uncertainty.
\label{fig:dhti}}
\end{center}
\end{figure}

Figure~\ref{fig:dhti} presents $\log$(Ti/H) against the (D/H)
values reported in the literature.  The visual impression is suggestive
of a correlation between the two quantities.  The non-parametric
Spearman and Kendall
correlation tests reject the null hypothesis at the 96$\%$ and 95$\%$ c.l.\
respectively and indicate a positive correlation between the two 
quantities.  
This correlation provides preliminary
evidence that the variation in Galactic D/H is related to 
the physical conditions that imply high depletion levels.  
Indeed, the results are consistent with the interpretation that low
Galactic D/H values are due to an increased depletion of D onto grains.

Examining the figure, it is obvious that additional
measurements of sightlines with large (D/H) value will be especially
valuable for testing the correlation.
Furthermore, one observes a 
large spread in Ti/H at low D/H.  If confirmed by future observations,
the spread in Ti/H would argue against a simple linear correlation
between Ti/H and D/H.
We note that two additional stars ($\zeta$\,Pup, $\gamma^2$\,Vel) with
(D/H)~$> 1\sci{-5}$ have \ion{Ti}{2} column densities reported
in the literature \citep{welsh97}.  The inferred Ti/H value for
$\zeta$\,Pup ($\log$(Ti/H) = $-8.5 \pm 0.08$)
follows the trend indicated in Figure~\ref{fig:dhti}.
The upper limit to Ti/H $(<-8.9$)
for $\gamma^2$~Vel, however, suggests the gas 
along this sightline is significantly depleted even though its
D/H value is among the highest known.  A similar conclusion may be drawn
from the GHRS observations of \ion{Fe}{2} by \cite{fitzpatrick94} 
although line-saturation is a potential concern.  
Although the confirmation of 
a high depletion level toward $\gamma^2$\,Vel would 
raise concern,  \cite{draine04a} notes that the processes of D depletion
are different from those for other ions, e.g., D may be depleted
in polycyclic aromatic hydrocarbons
while heavier elements are depleted by other types
of dust grains.  Therefore one need not
expect a one-to-one correspondence between Ti/H and D/H. 
In the coming year, we plan to acquire \ion{Ti}{2}
observations along many additional sightlines to 
better constrain the slope and scatter of Ti/H vs.\ D/H.

We have also examined the correlation between Ti/H and the 
$\N{HI}$ values of the sightlines.  Because one observes
a correlation between volume density and depletion level
\citep[e.g.][]{jenkins87}, one may expect a similar trend
for $\N{HI}$.  
Furthermore, if photoionization is important along these sightlines 
its effects should correlate with $\N{HI}$.
The Spearman and Kendall tests,
however, reveal the null hypothesis is ruled out at only the
$29\%$ and $45\%$ c.l.\ respectively.
This follows the results from previous surveys of the Galactic
disk \ion{Ti}{2} \citep{stokes78,welsh97} although \cite{wakker00} see a strong
correlation in a set of clouds with much larger dynamic range in $\N{HI}$
than the sightlines considered here.

\begin{figure}[ht]
\begin{center}
\includegraphics[width=3.6in]{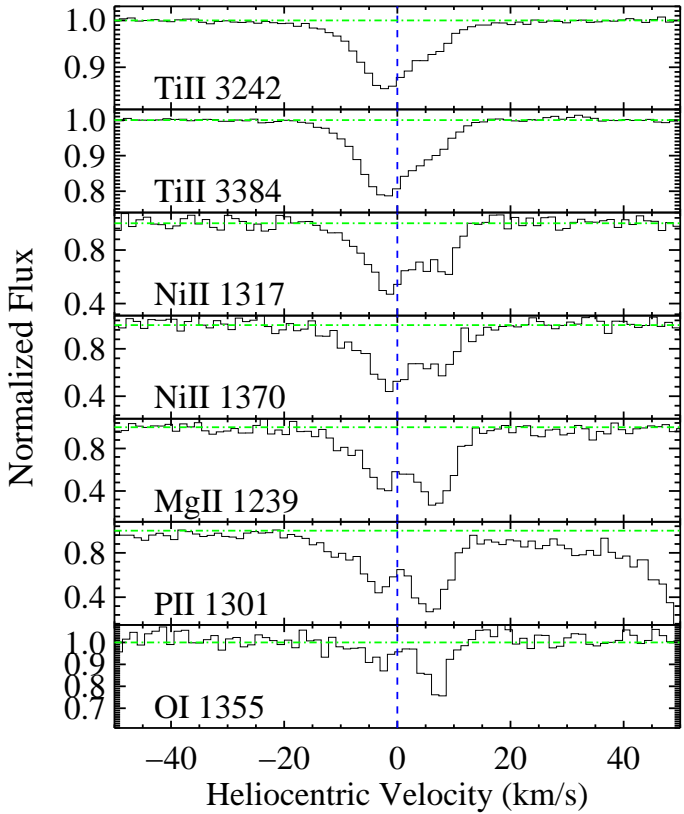}
\caption{Velocity profiles for the ISM sightline toward HD~195965.
Separately, the refractory ions (Ni$^+$, Ti$^+$)  and non-refractory
ions (P$^+$, O$^0$) show very similar sets of profiles.  One notes,
however, that the positive velocity component has a much larger
depletion level for Ni$^+$ and Ti$^+$
($\approx 10\times$) than the negative component.
}
\label{fig:compion}
\end{center}
\end{figure}

We wish to emphasize an important aspect
of differential depletion in the analysis of Galactic D/H.  
In Figure~\ref{fig:compion},
we present the \ion{Ti}{2} profiles for the HD~195965 sightline against a
series of UV transitions obtained with STIS on the {\it Hubble 
Space Telescope}.  The UV data have higher resolution (FWHM~$\approx 1.5$\kms)
but lower S/N per resolution element than the optical observations.  
Examining the detailed
component structure of the profiles, the refractory
ions (e.g.\ Ti$^+$, Ni$^+$) have similar characteristics. 
Comparing against the non-refractory profiles (P$^+$, O$^0$), however,
we identify the two main components but note that the relative
abundances are significantly different.  The Ti abundance relative to O
is $10\times$ greater in the positive component
than the negative component. 
It is also noteworthy that the \ion{Mg}{2} profiles more closely
track the non-refractory species even though Mg is depleted
along the sightline.

This raises a number of concerns regarding the analysis 
and interpretation of D/H.   
First, variations in the depletion level on a given sightline would generally
lead to a weaker correlation between any existing Ti/H vs.\ D/H
correlation if one only considers the integrated values.
By a similar token, one would tend to underestimate the magnitude of
intrinsic scatter in D/H regardless of the physical mechanism 
responsible.
Another complication is that it is unclear whether one should 
constrain the analysis
of D in lower resolution data (e.g.\ FUSE observations) with the 
velocity profiles traced by the \ion{O}{1}, \ion{Mg}{2}, or \ion{Ti}{2}
profiles.  Profiles of \ion{Fe}{2} are likely to be similar
to \ion{Ni}{2} and \ion{Ti}{2} and could also be confusing in analysis
of \ion{D}{1}.
A useful exercise would be to quantitatively compare 
high resolution D profiles with both refractory and non-refractory
transitions along sightlines where the depletion levels vary.
We consider these issues 
and examine the systematic effects on the inferred D/H ratios 
in a future paper.
Finally, we note that the D profile of at least
one extragalactic sightline shows significantly different component
structure than the corresponding metal-line profiles \citep{tytler96}.  
Although the physical origin is more likely related to differences
in metallicity or the ionization state of the metals, one cannot
rule out intrinsic scatter in D/H even in these low metallicity
sightlines.  This issue could be relevant in studies of
D/H in damped \lya\ systems.

\acknowledgments
We acknowledge the efforts of S. Vogt, G. Hill, and the HIRES upgrade
team for their efforts in completing the CCD mosaic upgrade.
We would like to thank the referee B. Draine for helpful comments
and criticism.  Similarly, we appreciate the input from E. Jenkins,
W. Moos, C. Oliveira, and K. Sembach.
JXP is partially supported by NSF grant AST-0307408. 
TMT appreciates support from NASA through grant NNG04GG73G.
This research was supported in part by the National Science Foundation 
under Grant No.\ PHY99-07949.

\end{document}